# Pulsar Positioning System:
## A quest for evidence of extraterrestrial engineering

Clément Vidal
Center Leo Apostel
Vrije Universiteit Brussel
Krijgskundestraat 33,
1160 Brussels, Belgium
http://www.clemvidal.com,
contact@clemvidal.com

**Abstract**: Pulsars have at least two impressive applications. First, they can be used as highly accurate clocks, comparable in stability to atomic clocks; secondly, a small subset of pulsars, millisecond X-ray pulsars, provide all the necessary ingredients for a passive galactic positioning system. This is known in astronautics as X-ray pulsar-based navigation (XNAV). XNAV is comparable to GPS, except that it operates on a galactic scale. I propose a SETI-XNAV research program to test the hypothesis that this pulsar positioning system might be an instance of galactic-scale engineering by extraterrestrial beings (section 4). The paper starts by exposing the basics of pulsar navigation (section 2), continues with a critique of the rejection of the extraterrestrial hypothesis when pulsars were first discovered (section 3). The core section 4 proposes lines of inquiry for SETI-XNAV, related to: the pulsar distribution and power in the galaxy; their population; their evolution; possible pulse synchronizations; pulsar usability when navigating near the speed of light; decoding galactic coordinates; directed panspermia; and information content in pulses. Even if pulsars are natural, they are likely to be used as standards by ETIs in the galaxy (section 5). I discuss possible objections and potential benefits for humanity, whether the research program succeeds or not (section 6).

## Contents

# 1   Introduction

Navigation is a universal problem whenever one needs to go from point A to point B. Around Earth's orbit, Global Navigation Satellite Systems (GNSSs) such as the American Global Positioning System (GPS) or the Russian GLObal NAvigation Satellite System (GLONASS) have revolutionized the way planes, boats, cars and pedestrians navigate. GPS did not appear overnight, but was the culminating outcome of ground-based radio navigation in the 1920s, gradually complemented by satellites. These satellites broadcast timing information that allow users to determine accurately not only their instantaneous position, but also their instantaneous velocity. GPS requires at least 24 satellites equipped with precise atomic clocks and algorithms calculating the position of satellites, which must correct for relativistic effects predicted by Einstein's theory. GNSSs constitute a great achievement of modern science and engineering, and will continue to prove revolutionary for all kinds of location-based services in our Internet era. However useful on Earth, GNSSs are of little use for space missions in the solar system and in deep space. We could only dream of the equivalent of a GNSS on a galactic scale. Remarkably, though, it seems already to exist.

Back in 1972, Carl Sagan, with Linda Sagan and Frank Drake, famously composed "A message from Earth", to be placed on Pioneer 10 as a way of communicating our position in the galaxy to any extraterrestrials who happened upon the probe. The spacetime coordinates of Earth are encoded in the message, thanks to a reference to 14 pulsars and the galactic center (Sagan, Sagan, and Drake 1972, see figure 1).

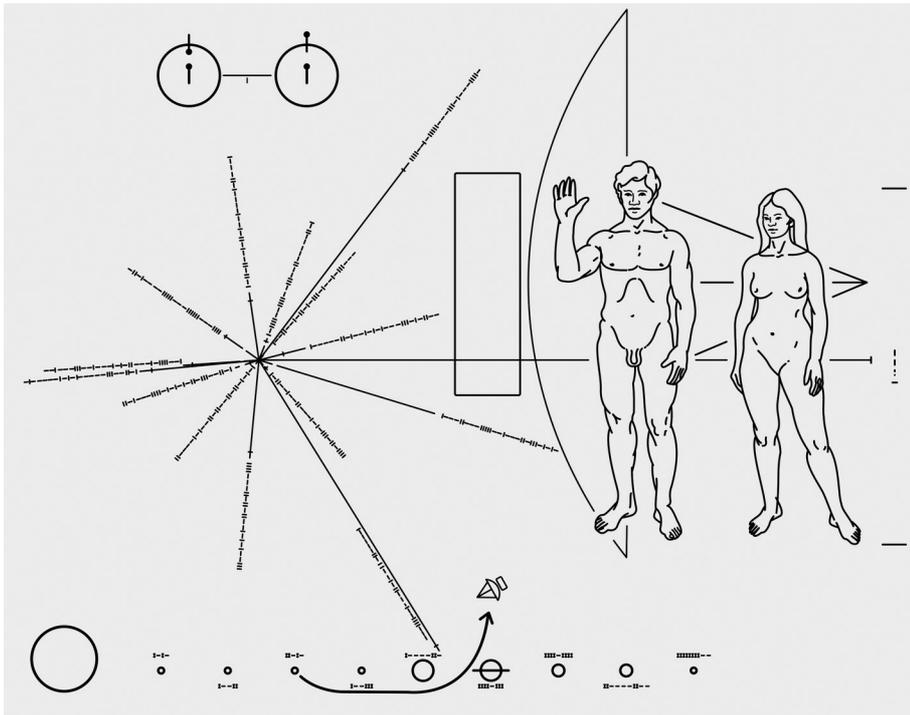

Figure 1: The Pioneer 10 plaque. On the left, the position of the Sun is shown relative to 14 pulsars and the center of the galaxy

However, the potential of *radio pulsars* for galactic positioning and navigation was fully explored only two years later by Downs (1974). At that time, Downs showed that spacecraft position could be determined with an error on the order of 1500 km in the solar system, and he pointed out ways to narrow down the error.

Achieving an accuracy of even a thousand kilometers may seem largely sufficient on a galactic scale, but it is not enough for solar-system missions. For example, in order to land a spacecraft on a particular crater of a planet, higher accuracy is needed. The technique of pulsar-based navigation took a leap forward with the suggestion of Chester and Butman (1981) to use *X-ray pulsars* instead of radio pulsars. With their methodology, they showed that an accuracy of 150 km could be achieved. One important advantage of X-rays over radio waves is their short wavelength, which means they can be detected with small detectors that are easy to engineer into a spacecraft.

One year after, the first millisecond pulsar (MSP) was discovered in the radio band (Backer et al. 1982) and in 1993, X-ray emissions from a millisecond pulsar were discovered (Becker and Trümper 1993). These discoveries changed pulsar navigation, as MSPs have stabilities comparable to atomic clocks used in GNSS satellites.

For both GNSSs and pulsars, the accuracy is limited by the stability of the clocks employed. Typically, GNSSs have clocks with an accuracy of 3 nanoseconds (ns), while the interstellar medium leaves measurement uncertainties of the order of 100 ns for millisecond pulsars (Verbiest et al. 2009). Multiplied by the speed of light, this translates into errors of 90 cm and 30 m respectively. This means that navigation with X-ray millisecond pulsars has the potential to be accurate down to 30 meters on a galactic scale. Engineering and observational constraints make it hard to achieve such small error margins, but recent methods suggest that a range error of about 100 m may be achieved (e.g. Sheikh, Golshan, and Pines 2007; Hanson et al. 2008).

Today, X-ray Pulsar Navigation (XNAV) is an active field of research and engineering (see e.g. Sheikh 2005; Sheikh et al. 2006; Becker, Bernhardt, and Jessner 2013). There is a growing international space-navigation interest in XNAV. For example, on November 9[th] 2016, the China Academy of Launch Vehicle Technology successfully launched the world's first mini-satellite to test autonomous spacecraft navigation in space (see e.g. Spaceflight101 2017). The European Space agency is also interested in the concept (e.g. Sala et al. 2004), as well as the American NASA Station Explorer for X-ray Timing and Navigation Technology (SEXTANT) mission, which also plans to demonstrate real-time on-board X-ray Pulsar Navigation (Winternitz et al. 2016).

To sum up, pulsars can be used as timekeeping devices comparable to the atomic clocks of the 1990s, and can be used as a galactic positioning system comparable in functionality and accuracy to today's GNSSs.

Did the suitable set of pulsars for galactic navigation appear by chance? Or could it have been set up in part or in totality by extraterrestrial intelligence (ETI)?

Even if no ETI engineering is involved, what are the SETI implications of the existing timing and navigation capabilities of pulsars?

In this paper, we study the possibility that extraterrestrials indeed set up, or instead are simply making use of a *Pulsar Positioning System* (PPS). If successful, proving that a PPS was engineered would constitute a proof of extraterrestrial intelligence. If unsuccessful, even a natural galactic timing and navigation solution has profound consequences, as it may provide a common communication standard and ground for all civilizations in the galaxy. The study of XNAV and the testing of SETI-XNAV also promises to advance our knowledge of positioning and navigation systems for solar system and deep space missions, and may also impact the design of future GNSSs for planet Earth.

## 2 Pulsar navigation

Before introducing the basics of pulsar navigation, I present here the distinction between normal and millisecond pulsars, and then report some of the rich behavior of pulsars.

### 2.1 Normal and millisecond pulsars

An essential observation about pulsars is that there are two distinct populations: normal pulsars, and millisecond pulsars (MSPs). "Normal" pulsars have a pulse period $P \sim 0.5$ s, while MSPs have a period $P$ between 1.4 ms and 30 ms. The two populations are usually illustrated with a $P - \dot{P}$ diagram (see Fig. 2), where $\dot{P}$ denotes the period derivative, i.e. the rate of change in the pulsation period. The smaller $\dot{P}$ is, the more stable the period is.

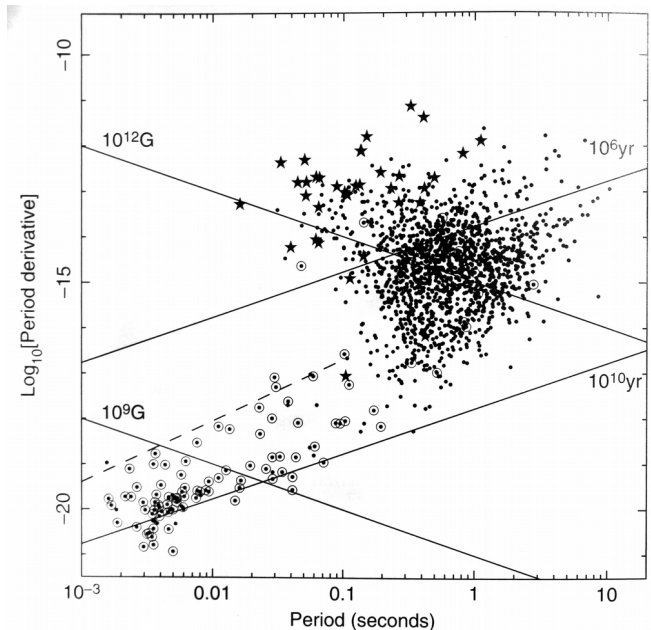

Figure 2 - $P - \dot{P}$ diagram showing two populations of pulsars: normal pulsars in the upper-right, and millisecond pulsars in the lower-left. Binary pulsars are shown as circles, normal pulsars as points, and young pulsars associated with supernova remnants as stars. Adapted from (Lyne and Graham-Smith 2012, 152)

Let us highlight a few of the key differences between normal pulsars and MSPs (summarized in Table 1). Regarding their origin, normal pulsars are the remains of the collapse of stars more massive than 10 solar masses, while the origin of MSPs is tied with binary stars. Their history can be more complex, but the standard model (Bhattacharya and van den Heuvel 1991) says that, in a binary star system, a neutron star is formed during the supernova explosion of the massive star. If the other companion star is massive enough, it will evolve into a giant and overflow its Roche lobe, towards the neutron star. The neutron star can then gain a spin boost by accreting matter from the companion star. Such pulsars are often called "recycled" or "rejuvenated" pulsars, because they gain a new vitality by accreting their companion star. MSPs are thus mostly found in binary systems. An entry point to the literature about pulsar and MSP formation scenarios can be found in (Lorimer 2008).

MSPs represent about 10% of the total known pulsar population. They are distributed isotropically in the galaxy, by contrast with normal pulsars that are more concentrated in the galactic disk. This may be an important property for navigation purposes, if the observation is not entirely due to selection effects (Lorimer and Kramer 2005, 26). MSPs are ~100 000 times more stable than normal pulsars. The magnetic field of MSPs stays at ~$10^8$ G, which is orders of magnitude less than normal pulsars. MSPs have a lower velocity than normal pulsars. Regarding the pulse shape, MSPs display on average one more component than normal pulsars. A component is defined as a Gaussian curve appearing in the average pulse profile. Some MSPs have complex profiles displaying 9 or even 12 components – respectively J1012-5307 and J0437-4715 (Kramer et al. 1998, 277).

| Property | Millisecond Pulsars | Normal Pulsars | References and notes |
|----------|---------------------|----------------|----------------------|
| Percentage of the observed population | 10% | 90% | See e.g. (Becker, Bernhardt, and Jessner 2013). Selection effects may play a role (see Lyne et al. 1998 for details). |
| Binarity | 80% | 1% | 80% of MSPs have a companion star, while only 1% of normal pulsars do (Lorimer and Kramer 2005, 27). |
| Galactic distribution | Isotropic | Towards the galactic plane | The distribution of normal pulsars is towards the galactic plane, while MSPs' is more isotropic (Lyne and Graham-Smith 2012, 105). |
| Period | ~0.003s | ~0.5s | See e.g. (Lorimer and Kramer 2005, 26) |
| Period derivative (stability) | ~$10^{-20}$s.s$^{-1}$ | ~$10^{-15}$s.s$^{-1}$ | MSPs are comparable in stability to atomic clocks of the 1990s (see e.g. Lorimer and Kramer 2005, 26). |
| Magnetic field | $10^8$-$10^9$ | $10^{12}$ | (Lyne and Graham-Smith 2012, 168). |
| Velocity | Low | High | MSPs have a velocity 100km.s$^{-1}$ lower than normal pulsars (Lyne et al. 1998; Hobbs et al. 2005). |
| Number of pulse components | $4.2 \pm 0.4$ | $3.0 \pm 0.1$ | These are average values (Kramer et al. 1998, 277). |
| Timing noise | Rare | Common | (Lyne and Graham-Smith 2012, 89) |
| Glitches | Exceptional | Rare | (Espinoza et al. 2011). Exceptions of MSP glitches are PSR B1821−24 (Cognard and Backer 2004) and J0613−0200 (McKee et al. 2016). |

Table 1: Comparison of some properties of normal and millisecond pulsars. More distinguishing features can be found in the literature (e.g. Kramer et al. 1998).

When observing pulsars for longer than days, two kinds of timing irregularities appear. First, the *timing noise*, which is a general erratic behavior; and *glitches*, which are abrupt changes in the rotation speed. Both these timing irregularities are reduced in the case of MSPs.

To sum up, MSPs have unique timing features: notably short periods, high stability, and few timing irregularities. Their low velocity and isotropic distribution in the galaxy makes them especially suitable as navigation beacons. The population of MSPs thus possesses features that distinguish them normal pulsars, and many of these features are useful for the task of navigation.

## 2.2 Pulsar behavior

*A visitor to a pulsar observing session will see on the oscillograph something quite unlike anything in the rest of astrophysics: a never-ending dancing pattern of pulses: sometimes bright, sometimes faint, sometimes in regular patterns, sometimes disordered, sometimes switching off entirely only to resurge with greater vigour. Variations can be found on every time scale down to tiny fractions of seconds.*

*Astrophysics is a field used to dealing with objects which evolve over millions, over thousands of millions of years, perhaps occasionally punctuated by dramatic cataclysmic events, but generally affording no more than an unvarying image through the telescope. How are we then to deal with a phenomenon which is so alien to the common astrophysical experience?*

As Rankin and Wright (2003, 43–44) explain, pulsars stand out in the landscape of astrophysical objects.

The initial lighthouse model (Pacini 1967; Gold 1968), however useful, is inadequate to account for the rich behavior of pulsars, gradually uncovered by decades of modern astronomy and astrophysics. Instead, pulsar literature contains a variety of models that deal with the variety of phenomena exhibited by pulsars, explained sometimes only partially.

We can distinguish four scales of pulses (Lyne and Graham-Smith 2012, 230). The *average pulse profile* is an average of hundreds of *pulses*. It is remarkably stable over time, although every single pulse is different (see Figure 3). This stability is of key importance in pulsar timing measurements (Lorimer 2008). *Sub-pulses* appear inside pulses and last about 1 μs, while one speaks of *microstructure* when describing what appears inside subpulses (about 1 ns).

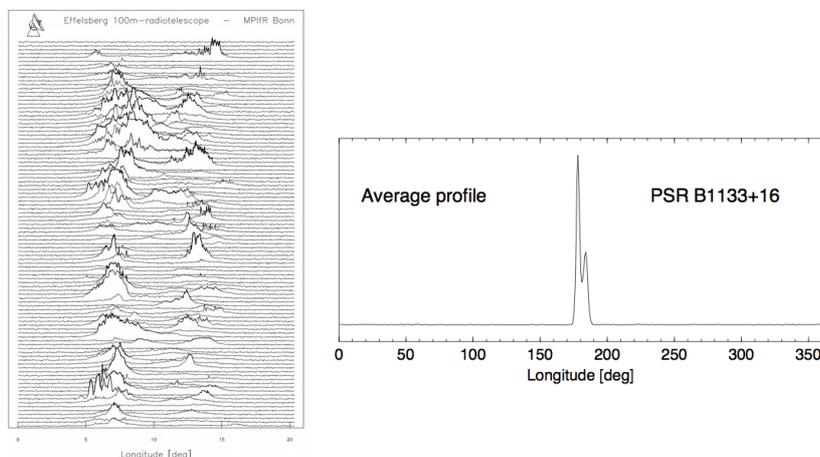

Figure 3 - Individual pulses of pulsar PSR B1133+16 vary in shapes and strength (left), whereas the average profile is stable (right). Adapted from (Kramer 2004).

Some pulsars also display *nulling* and *mode changing* (see e.g. Manchester 2009). Pulsar *nulling* appears mostly in long period pulsars, where the emission abruptly turns off for a duration varying from a few pulses to weeks, and then suddenly turns back on. *Mode changing* appears when the average pulse profile abruptly changes to a different shape and then reverts to its original shape. Nulling and mode changing are arguably related, since they both occur in older pulsars and on similar timescales (Lyne and Graham-Smith 2012, 218). Still, they are not easy to model and explain. For example, the nulls in PSR B1133+16 are not simultaneous across all frequencies, suggesting that it is not simply the pulsar emission turning off (Bhat et al. 2007). Exceptionally, pulsar nulling also appeared in a MSP (see PSR J1744-3922 in Faulkner et al. 2004).

Another phenomenon is a sudden increase in rotation speed of a pulsar, or *glitch*. Generally, MSPs don't glitch. However, a MSP (PSR B1821−24) did by a rotational frequency of 3 nHz (Cognard and Backer 2004), and another (J0613−0200) glitched by 0.82 nHz (McKee et al. 2016)!

Also remarkable is the case of accreting millisecond pulsars that *spin down*, although we mentioned that the standard model says accretion should *spin up* and not spin down neutron stars. A limit case is MSP SAX J1808.4–3658 displaying spin up and spin down phases (Di Salvo et al. 2008). More details on the rich phenomenology of accreting millisecond pulsars can be found in the work of Lamb and colleagues (Lamb et al. 2009).

The pulse profile of MSPs can also change depending on the frequency, a phenomenon not observed with normal pulsars. For example, PSR J2215+5135 "shows a double pulse profile at 350 MHz, but by 2 GHz is predominantly single peaked, which peak is neither of the peaks seen at 350 MHz" (Roberts 2012).

To sum up this cursory overview of pulsar behavior, one might observe that unified models do not exist, and may not even be possible.

## 2.3 Navigation with pulsars

*The unique properties of pulsars make clear already today that such a navigation system will have its application in future astronautics.*

(Becker, Bernhardt, and Jessner 2013)

As Becker, Bernhardt and Jessner note, pulsars are uniquely suited to constitute a galactic navigation system. Let us first see why by surveying some alternatives.

For space navigation, classical methods use Earth-based stations to guide spacecraft, but give rise to increasing uncertainties as spacecraft travel further from the Earth. The uncertainty is ±200 km at the orbit of Pluto and ±500 km at the distance of Voyager 1 (Becker, Bernhardt, and Jessner 2013, 1). Other limitations include communication delays, and the weakening of signals at large distances.

Why not use basic geometry? At first, space navigation may not seem so difficult, provided several stars can be observed and their relative angle calculated. However, because the angular resolution of sensors is not sensitive enough, this method leads to uncertainties of several thousand kilometers (Becker, Bernhardt, and Jessner 2013, 2).

Why should variable sources be used, and not static ones? The measurement process and position calculation takes time, and each time one wants to check the position, new measurements and new calculations need to be done. With static sources, position and velocity are not computable in real-time or in a precise manner. Already back in 1972, Sagan *et al* (1972, 883c) also rejected star map position indicators, because of "stellar proper motions and serious data-handling problems in decoding". By contrast, a variable source such as a pulsar "provides a periodic signal that assists in the prompt identification of each specific source, since most of these signatures are of unique period and strength" (Sheikh 2005, 58).

Why not use variable *visible* sources? The issue here is that there are too many of them, and such plurality makes their identification more difficult (Sheikh 2005, 66). However, Sheikh also argues that in the *visible* or *gamma-ray* band, there are too few pulsars for the task of navigation.

Why not use *radio* pulsars? After all, pulsars were discovered in the radio band and that is the wavelength in which they are most studied. The practical issue with regard to this solution is that radio waves have long wavelengths so that radio antennas with 20 m dishes in diameter (or larger) must be used in order to discriminate the signal (Emadzadeh and Speyer 2011, 10). Of course, the inclusion of such large antennas would greatly affect the cost and feasibility of spacecraft designs and space missions.

Why millisecond X-ray pulsars (MSXPs)? Not all pulsars are suitable for navigation, and MSXPs are presently the best choice. To sum up, to determine time, position and velocity, it makes most sense to make use of sources with the following characteristics (Sheikh 2005, 34):

1. They should be intense, so small detectors can detect them.
2. They should have stable, periodic signatures, so the behavior is highly predictable with simple models, placing a low computational load on the spacecraft.
3. They should exhibit narrow and sharp pulse profiles, so arriving pulses are quick and easy to identify.
4. They should have a unique signature, such as a pulsar's unique average pulse profile (APP). Like a fingerprint, this uniqueness of APPs avoids confusion with other sources.
5. These sources need to be highly stable, to determine position, velocity and attitude with high accuracy.
6. The sources need to comfortably penetrate the interstellar medium, so that the signal is robust to common galactic interference. X-rays can typically go through the interstellar medium.

MSXPs meet all these desiderata. However, almost all MSXPs are in binary systems (Lyne and Graham-Smith 2012, 151) and one may think that the binary motion would complicate the pulsar timing. Fortunately, this can easily be corrected (Blandford and Teukolsky 1976).

The properties of X-ray pulsars makes them ideal as navigation beacons, where an individual pulsar plays a similar role as a GPS satellite. Both GNSSs and XNAV use a navigation method based on *Time of Arrival* (ToA). This method consists in comparing the pulses' ToA measured, with a predicted ToA at a reference location. By observing three pulsars or more, a position can be determined (see Figure 4).

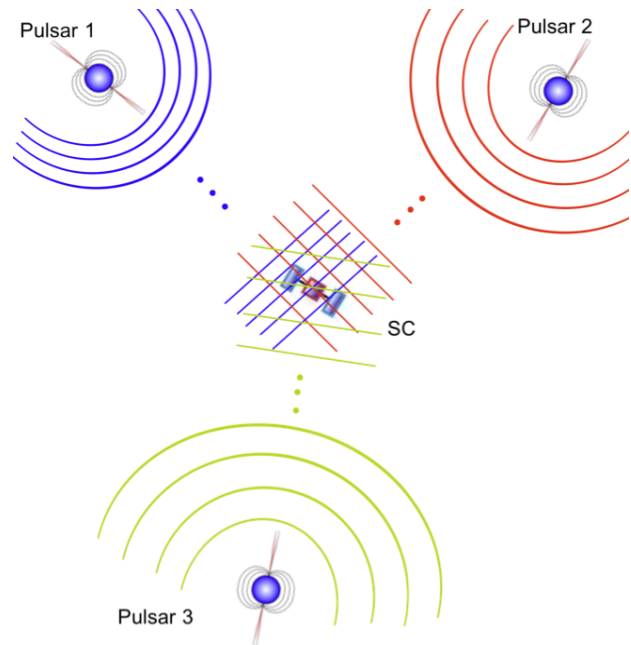

Figure 4: A three dimensional position fix can be obtained by observing at least three pulsars. Given three well chosen pulsars, there is only one unique set of pulses that solves the location of the spacecraft (SC). Figure adapted from (Sheikh 2005, 200).

In navigation science, there is a distinction between *absolute positioning*, answering the question "Where am I?", and relative *positioning*, answering the question "Where am I now in relation to where I was before, or in relation to another external navigation aid?". Both GNSSs and XNAV address and solve the absolute positioning issue. For XNAV, this requires the navigation system to compare measured and predicted pulse times of arrival. With a pulsar almanac and a timing model, the spacecraft can identify the different pulsar sources, from which its exact location can be determined (for details about the methodology and algorithms, see Sheikh 2005, chap. 6). This means that the GNSS or spacecraft user doesn't need to keep track of its position relative to its point of departure to compute its position. The spacecraft doesn't need external assistance, and we therefore speak about *autonomous navigation*.

Generally, the longer the pulsars are observed, the more accurate the positioning can be determined. For example, an accuracy of hundreds of kms is achieved with a short integration time; but can be reduced to about 100 m with more than one week of integration time (Buist et al. 2011, 158a).

    If we compare GNSS and pulsar navigation, they are similar in many respects (see Table 2). They are both highly accurate navigation systems based on ToA of signals. Both pulsars and GNSS satellites broadcast a regular signal for unambiguous identification (respectively the average pulse profile and the coarse/acquisition code). They are both passive navigation solutions, because no transmission between the user and the satellite/pulsar segment is needed (see Figure 7). However, GNSS satellites broadcast their time and position, while pulsar distances must be evaluated with astronomical observations.

    To sum up, a pulsar positioning system (PPS) allows one to determine *instantaneous position, instantaneous velocity* (by taking into account the Doppler compression or expansion of pulsar signals) and *attitude* (i.e. determination of the spacecraft orientation). It allows one to easily correct relativistic effects, to solve the "lost in space" or "cold-start" situation, i.e. the determination of position and velocity without requiring external assistance (Sheikh 2005, 51). Compared to Earth-based navigation solutions for space missions, the advantages of a pulsar positioning system are the following (J. Liu et al. 2010):

    1. it is entirely passive;
    2. its navigational accuracy does not decline with time, as constant correction is available from different X-ray sources;
    3. it is robust to interference;
    4. it seems fit for the whole outer space.

The last point (4) is not certain, but appears to be a reasonable assumption. How can we know if XNAV works everywhere in the galaxy? We only have access to the pulsar positioning system in a tiny part of the galaxy. As long as we are unable to travel to the four arms of the galaxy and test if XNAV is operational, it seems that we cannot know. However, modern astrophysics has methods to estimate the position and distribution of pulsars, and thus a first approximation of the XNAV coverage.

| Property | GNSS | XNAV | Notes |
|---|---|---|---|
| **Accuracy** | ~ 3 ns (1 m) | ~ 100 ns (30 m) | The accuracy of pulsars may be improved with longer integration times or better observational techniques |
| **Navigation method** | Time-of-Arrival-based | Time-of-Arrival-based | The two navigation methods are similar and use similar algorithms to determine position, velocity and attitude |
| **Regular signal for beacon identification** | Coarse/Acquisition (C/A)-code | Stable pulse profile | The C/A-code is constantly repeated in the transmitted GPS signal. See eg. (Buist et al. 2011, 156a). This property is similar to the average pulse profile of pulsars, which is stable. |
| **Passive navigation** | Yes | Yes | No transmission between the user and satellite/pulsar segment is needed |
| **Absolute navigation** | Yes | Yes | No tracking with the point of departure is needed |
| **Time and position transmission** | Yes | No | While GNSS satellites broadcast their time and position, pulsar time needs to be parameterized and distances must be evaluated with astronomical observations |

Table 2 - Satellite and pulsar navigation are similar in many respects.

To suppose that XNAV would work only in our small solar-system region would be a blatant violation of the Copernican principle. Following this principle, we must assume that Earth has no special position, and that XNAV is available in most parts of the galaxy. Said otherwise, anthropocentric hypotheses are needed to make consistent the idea that XNAV would work only near our solar system.

Today, PPS applications are mostly foreseen for autonomous navigation in the solar system. Looking towards the deep future, the applications of a PPS for the galaxy will be as rich and profound as GPS was for planet Earth.

# 3   Dismissing the dismiss

*A major difficulty with the pulsar discovery*
*was its suspiciously artificial nature.*
*All the facts pointed to a celestial origin,*
*but celestial and artificial imply...aliens.*

(McNamara 2008, 46)

The discovery of pulsars surprised astronomers. The unusual high frequency of the pulsations as well as their regularity immediately led to the hypothesis that it could be ETI. However, the ETI hypothesis was quickly dismissed because astrophysicists developed instead a natural "lighthouse model", in which rotating neutron stars are at the origin of the pulsations (Pacini 1967; Gold 1968).

Traditionally, the story of the discovery of pulsars is told as a case where a signal is wrongly assumed to be an artificial one. Artificial signals have been attributed to pulsars at least two times. When Jocelyn Bell reported the strange signal to Antony Hewish in 1967, they thought the signal was man-made (Wade 1975, 360c; Hewish 2008). After the human interference hypothesis was dismissed, the alien intelligence interpretation was taken seriously in the earlier stages of the discovery, and pulsars were nicknamed "LGM" stars, for "Little Green Men" (Penny 2013, 4).

The second time was in 1989, when a team of astronomers first discovered rapid pulses coming from the pulsar remnant of supernova 1987A (Kristian et al. 1989). Another set of rapid pulses was later discovered. The astronomers assumed that it was again the pulsar...whereas it turned out to be due to human interference (Anderson 1990)!

To sum up, based on a superficial analysis of the signal, it is not obvious to tell if we are dealing with "Little Green Men" or little human interferences. Even today, specialists reflect on how to distinguish a pulsar from an artificial beacon (J. Benford 2010).

Historically, the pulsar extraterrestrial hypothesis was dismissed on five grounds:

(1) the energy expanded was too vast to be ETI,
(e.g. Jastrow and Thompson 1977)
(2) other sources of pulsation were quickly found (Burnell 1977),
(3) the source didn't come from a planet and thus could not come from an intelligent civilization (Hewish 2008),
(4) the source was not narrow-band (Hewish 2008),
(5) there exists a natural model explaining pulsars
(Pacini 1967; Gold 1968).

Let us critically discuss each of these reasons.

## 3.1   Too much energy

Regarding (1), the amount of energy radiated by a pulsar is indeed impressive… by Earth standards. But there is no sense to be geocentric when making assumptions about ETI energy use. If the total energy of the Galaxy is ~ 4 x $10^{44}$ erg.s$^{-1}$ (Kardashev 1964), and if a typical pulsar such as the Vela pulsar radiates ~ 6 x $10^{32}$ erg.s$^{-1}$, the known 2000 pulsars would use about 0.0000003 % of the total energy available in our galaxy. Setting up a network of pulsar beacons may be considered cheap and worthwhile for a galaxy-spanning civilization.

The galaxy is about 160 000 light years wide (see e.g. Xu et al. 2015), so at least for navigation, signals need to be visible from far away, and thus require a lot of energy. Furthermore, we saw that higher X-rays energies are useful to navigate with small spacecraft and to deal with the interstellar medium.

What cost-benefit analysis could justify such an expansive and expensive engineering? The usefulness of a common galactic timing and navigation solution could be such a justification. The applications of our own GNSSs on Earth go well beyond the initial military motivation to guide high-speed missiles. Accordingly, even a GNSS remains expensive to design, to launch and to maintain. So, a pulsar positioning system on a galactic scale would indeed cost orders of magnitudes more effort.

A subtler answer could be that ETI modulates the energy of existing pulsars. Instead of letting a pulsar lose its energy, ETI would use and modulate that energy for a more intelligent purpose. The ETI would not have to *produce* this energy, just to *channel* or to *modulate* it (see also section 6.4).

Imagine that you describe a GNSS to an isolated hunter-gatherer of an Amazonian tribe. Would he be willing to believe that humans managed to send satellites in orbit just for navigation purposes? Wouldn't it look like an impossible feat, and even if it was possible, a waste of energy? From the perspective of his navigation needs in the jungle, no space-based system is required. This would seem like a futile project, and a waste of an enormous amount of time and energy.

## 3.2   Not unique

The second reason (2) why the pulsar ETI hypothesis was dismissed is that other sources of pulsation were quickly found. As Jocelyn Bell Burnell (1977) argued: "It was very unlikely that two lots of little green men would both choose the same, improbable frequency, and the same time, to try signaling to the same planet Earth".

There are two different arguments in this sentence. The first says that two ETIs using the same frequency is unlikely. I fail to see any logic in this argument. Imagine an alien coming to Earth, observing two mobile phones emitting on similar frequencies, and concluding that this is not intelligent communication. Far from being improbable, one may conjecture that any ET life to have created such a system would have coordinated the frequencies involved.

Now, the second part of the sentence further assumes that little green men "try signaling to the same planet Earth". It thus assumes anthropocentrically that pulsar signals are specifically directed to Earth. Here, I completely agree that it is unlikely that ETIs in two different parts of the galaxy would be eager to communicate with us at the same time. However, this is no reason to dismiss the ETI hypothesis. SETI researchers distinguish many different kinds of signals beyond specifically intentional signals targeted to planet Earth, such as unintentional signals (leakage, spacecraft communication), general scans or omnidirectional beacons (Tarter 1992).

A general counterargument is that we are unlikely to find just one instance of ETI. Statistically, astrobiologists generally assume that if we discover one instance of ET life, we would quickly discover others, or at least have strong grounds to assume that the universe is teeming with life (e.g. Webb 2015, 289).

## 3.3  Not a planet

The third argument (3) against the ETI interpretation comes from the discovery that the source didn't originate on a planet, and therefore could not have derived from an intelligent civilization. This seemed to be a decisive refutation of the ETI interpretation for Hewish (2008). Indeed, before making the discovery of pulsars public, Hewish waited to have more data, to measure a Doppler shift, and therefore deduce if the source was from a planet or not. Before that, he "felt compelled to maintain a curtain of silence" (Penny 2013, 4). The measurement result showed that the source was not a planet, and this was a major reason to reject the ETI interpretation.

The assumption that life must start and stay on a planet makes sense for low-energy astrobiology, i.e. when we look for traces of microbes or biospheres. However, the assumption doesn't hold if we engage in high-energy astrobiology, or the search for advanced extraterrestrial life forms. Such a search requires us to expand the search targets, and the constraint that ETI can only be found on an Earth-like planet can be relaxed (see e.g. Dyson 1960; Sagan 1973, chap. 6; Corbet 1997; Davies 2010; Bradbury, Ćirković, and Dvorsky 2011; Vidal 2014, chap. 9).

## 3.4  Not narrow-band

The fourth argument has to do with the bandwidth. SETI researchers expect to find narrow-band signals (see e.g. Siemion et al. 2010). So when Hewish discovered that the source was narrow-band, he indeed thought the source could be intelligent (Penny 2013, 4). However, the phenomenon of dispersion makes a pulse seen at longer wavelengths arrive behind the same pulse emitted at shorter wavelengths. So, in fact, the narrow-band signal observed was an observational effect (Penny 2013, 6). Pulsars generally emit through a broad band of frequencies, from ~ 100 MHz to ~ 100 GHz (e.g. Lorimer and Kramer 2005, 82). As Jastrow and Thompson (1977, 198) comment, it "would be wasteful, purposeless, and unintelligent to diffuse the power of the transmitter over a broad band of frequencies. The only feasible way to transmit would be to concentrate all available power at one frequency, as we do on earth when we broadcast radio and television programs."

We can challenge this line of thinking, first by recognizing that searching for narrow-band signals is indeed historically important, as it was recommended since the Cyclops report in the 1970s (Oliver and Billingham 1971). However, if this decision to choose narrow-band signaling made sense when we had mostly radio and television, the growth in information exchanges changes the situation (see e.g. Shostak 1995). Today, the growth of wireless multi-user communication systems has given rise to *spread-spectrum communication* techniques. Spread-spectrum communication systems are robust against the threats of jamming, interception, and cause minor interference to other systems. They are useful for "suppressing interference, making secure communications difficult to detect and process, accommodating fading and multipath channels, and providing a multiple-access capability" (Torrieri 2011, vii). As Messerschmitt (2012) argues in detail, this evolution in communication engineering must be reflected in our SETI search strategies, and broadband communication is definitely a serious option.

In a mixed scenario where a pulsar's energy is modulated by ETI, the broadband nature could also be the remains of the pulsar's natural emission, while the intentional communication would be encoded in some part of the spectrum.

## 3.5   Natural model

The best argument that pulsars are natural is that there is a natural model (Pacini 1967; Gold 1968). The establishment of this model came in several steps. Chadwick had discovered the neutron in 1932, and in parallel, Landau was theorizing that a star collapsing would achieve enormous density, and even predicted an upper-mass limit for white dwarfs, independently and earlier than Chandrasekhar (McNamara 2008, 22). However, as McNamara (2008, 2) recalls, at that time "it was thought neutron stars would be completely invisible and hence no one bothered to look for them".

Especially remarkable is Pacini's *prediction* of pulsars, six months before they were announced by Hewish and Bell. In his 1967 paper, Pacini expected an excited neutron star to form as the remains of a supernova. He proposed to look for ways to observe neutron star energy loss through magnetic or rotational energy, although he was rather skeptical about this observational possibility.

If Pacini's and Gold's lighthouse model is a natural, astrophysical way to explain pulsar behavior, why should we even bother to consider the ETI hypothesis?

There are three answers to this question. First, we don't necessarily need to challenge classical pulsar formation scenarios to seriously consider SETI-XNAV. We can speculate at different levels (see Table 3), and levels 1-3 could easily accommodate that there is a purely physical and natural origin of pulsars.

| Level | Comment |
|---|---|
| 0 – Natural | All pulsars are natural. We're just lucky they provide stable clocks and an accurate navigation system |
| 1 – Pulsars as standards | All pulsars are natural, but ETIs use them for timing, positioning and navigation purposes. Communication is galacto-tagged, and time-stamped with a pulsar standard |
| 2 – Natural and alterable | Some ETIs have the technology and capability to jam, spoof or interfere with a natural pulsar positioning system |
| 3 – Artificial MSXP for navigation | Only a few millisecond X-ray pulsars have been modified by ETI for galactic navigation and timing purposes |
| 4 – Artificial MSXP for navigation and communication | Only a few MSXPs have been modified by ETI, for navigation, timing and communication purposes |
| 5 – Artificial pulsars | All pulsars are artificial. ETI build them, even the new ones, by intentionally triggering supernovas |
| 6 – Artificial pulsars for us | All pulsars are artificial. ETI build them, and they are currently sending us Earth-specific messages |

Table 3 - Seven levels of artificiality of pulsars, from the least to the most speculative. Level 0 represents the current scientific belief. Level 1 doesn't require many hypotheses, but still has far-reaching consequences that we will outline in section 5. Level 2 is interesting to consider, even for solar-system navigation. If in the near future we start to rely heavily on XNAV, we need to know its degree of stability and robustness. Even if the broadband nature of pulsars makes them hard to jam (Buist et al. 2011, 153b), it is prudent not to underestimate the potential of ETI civilizations potentially billions of years our senior. Level 3 and 4 reflect my own subjective suspicion, but of course remains to be assessed and tested. Level 5 seems more like a science-fiction plot, and level 6 is anthropocentric, although the idea of radio pulsars sending Earth-specific messages has been defended (LaViolette 2006).

Second, Pacini's and Gold's models were the very first modeling attempts. Pulsar astronomy has immensely progressed since then, and pulsars display a phenomenology that requires much more advanced models (see section 2.2). There is no single unified pulsar model that can explain all the variety of observations.

Third, nobody predicted that our galaxy would host some pulsars with pulsations rivaling atomic clocks in stability, or that their distribution would make them useful for an out-of-the-spiral galactic navigation system.

Beside these points, any astrobiologist wants to avoid the aliens of the gap fallacy. As with the God of the gaps fallacy, the temptation is to attribute some unexplained phenomenon to aliens (Vidal 2014, sec. 9.1; Wright et al. 2014, sec. 2.3). Of course, this is unwarranted and unscientific. But as Wright *et al.* remark: "the other extreme – assuming that all observations must be the result of purely natural phenomena [and] are *not* due to advanced technology – is itself patently logically invalid because it *assumes* that ETIs have no detectable effect on the Universe".

At most, ease of modeling is a negative heuristic (Rubtsov 1991, 307; Vidal 2014, 217), in the sense that something simple to model can be disqualified as an ETI candidate. If there was just one model of pulsars explaining their origin and rich phenomenology, we could just dismiss the suggestion that ETI has anything to do with it. However, as Rankin and Wright (2003) summarize:

> three decades of experience and history have shown that general pulsar theories—physical theories of pulsars attempting to deduce the behaviour of real pulsars from first principles—are incapable of yielding significant, specific, falsifiable expectations about the observed emission of an actual individual pulsar.

Pulsar astrophysics remains a well-researched and well-established discipline, and successful in many respects. However, *we don't necessarily need to contradict existing pulsar models to entertain the possibility that ETI might be involved*.

Any kind of life must obey the laws of physics, otherwise it's magic! Let us illustrate this point with a biological example. We can describe a human body by its basic physical properties. For example, a physicist can say that a human body weights about 70 kg and contains about $7 \times 10^{27}$ atoms. There are two points about this description.

First, it is correct, although it will stay a crude and limited description. As much as a biologist doesn't need to challenge these physical facts to model human body, an astrobiologist doesn't need to challenge pulsar models to search for ETI traces in pulsars. Second, we can describe *any system* in terms of physics, but this does not differentiate whether the system is living or non-living.

A way to frame the problem is to consider the specification hierarchy of cosmic evolution and development (see Figure 5). Salthe comments that "biology, for example, cannot transcend chemistry or physics; it can only appropriate them, reinterpret them, rearrange them, harness them".

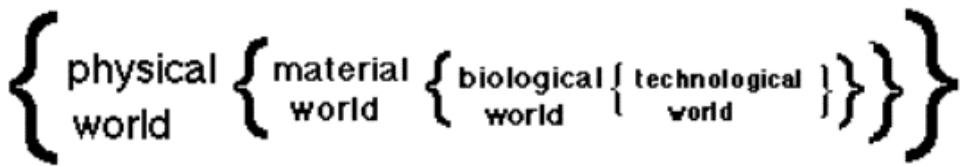

Figure 5 - A specification hierarchy showing stages in the development of the universe, with stages modeled as subclasses. From (Salthe 2002).

How can we tell if we are dealing with life or not? This is the tricky issue of defining life operationally and agreeing on biosignatures. I proposed an analysis of this delicate problem within the frameworks of thermodynamics and living systems theory (Vidal 2014, chap. 9; 2016a). Considering Figure 5, the challenge is to find behaviors belonging to the technological world that are not displayed by other subclasses. Yet, whatever the biosignature agreed upon, *the crux to gradually prove the existence of extraterrestrial life is to make and validate new predictions*.

The first step is thus to keep an open mind, and to give equal credence to the hypothesis that some astrophysical phenomenon may be purely physical as to the hypothesis that it may be alive or technological. The astrophysical or living model which explains and predicts the most should then be gradually favored.

Could it be that Jocelyn Bell's first intuition that she had discovered little green men was correct? Whatever will turn out to be true, intuitions and first impressions are no scientific method. Science is as much a work of observation and discovery, as it is of theorizing and (re)interpretation (Dick 2013).

## 4    The SETI-XNAV quest

*You've never encountered the questions,*
*let alone considering the answers*

(Smith 1957, 54)

Ten years before the discovery of pulsars, in his science fiction novel *Troubled Star*, George O. Smith pictures an extraterrestrial civilization wanting to make the Sun a variable star for the purpose of galactic navigation. The alien representative struggles to explain this need to humans. We are in a similar situation with galactic navigation. Very few of us consider the questions, even less are considering the answers. Yet we must enter these waters to try gradually to make new predictions, and start a quest for evidence of galactic engineering.

A well-known fact about pulsars is that they are useful, for fundamental physics, astrophysics, gravitational wave detection, timekeeping and galactic navigation (for details and references, see Vidal 2017, sec. 4). Even if pulsar usefulness may be intriguing, this provides no evidence of ETI activity.

Consider Earth's magnetosphere: it is clearly natural, and allows for navigation on Earth with the use of compasses. In the same way, if pulsars are useful for navigation, it doesn't necessarily mean that they are the product of extraterrestrial engineering. Indeed, there are many ways to navigate using natural resources. Birds or humans helped with stars or compasses can navigate planet Earth.

However, a compass provides only direction and will be biased towards the poles. It does not provide time, position, real-time velocity, and absolute positioning, which only GNSSs or a PPS can deliver accurately. Navigation with normal stars is also much less accurate than ToA navigation methods used in GNSSs and XNAV.

Additionally, not any pulsar will do when it comes to galactic positioning. We saw that *radio pulsar* navigation was proposed back in 1974 but is insufficient. High timing accuracy, short integration time, light and small antennas, etc. are needed to make pulsar navigation effective, and this is why pulsar navigation had to wait for the discovery of enough *X-ray millisecond pulsars* to become of practical interest.

The famous dictum "extraordinary claims require extraordinary evidence" (Sagan 1979, 73) is of no help here (see Schick 2002 for a detailed critique). To some, a navigation system comparable in accuracy to GPS, yet operating on a galactic scale may look extraordinary, but to most astrophysicists today, this looks normal and natural.

To illustrate the epistemic difficulty of the situation, imagine a smart hunter-gatherer who would have deciphered the electromagnetic waves sent by GPS satellites and found that they could be used as a navigation system. How could he argue or prove to his fellow tribe members that this reflects an intelligently set-up navigation system? The challenge is demanding, and we might be in a similar situation with the PPS.

A robust epistemological methodology that scientists use is one of conjectures and refutations (Popper 1959; 1962). To research whether pulsar navigation may have anything to do with ETI, we need to make the conjecture that ETI is involved with a PPS, and try to derive lines of inquiry leading to observable and testable predictions.

Let us now start such a SETI-XNAV quest, and see how it may lead to observable and refutable predictions that differ decisively from any purely physical models. Note that in what follows I gather the physical, natural, non-living assumptions under the "*astrophysical*" label, and the intelligent, artificial or living assumptions under the "*astrobiological*" label.

## 4.1  Galactic distribution

What is the galactic distribution of MSPs? We would expect that MSPs have a different distribution in the galaxy from normal pulsars, or from a random distribution. As a matter of fact, we already saw that MSPs' distribution is isotropic, while normal pulsars are more concentrated in the galactic plane (Lorimer and Kramer 2005, 26; Lyne and Graham-Smith 2012, 105). What is the likeliness for this to happen with natural scenarios? Why would there be more binary star systems outside the galactic disk giving rise to MSPs? Or could it be just a selection effect (Lorimer and Kramer 2005, 26)?

From the astrobiological hypothesis (levels 3-5), MSPs should have a distribution that is suitable for galactic navigation, significantly more than a random distribution, or than a distribution resulting from astrophysical models. There should thus be few redundancies in the coverage, and it should be better than what a random distribution may provide.

In terms of population, astrophysical models predict that the MSP population should be roughly equal to the normal pulsar population (Lyne et al. 1998). It should be different in the astrobiological view, where MSP population should be about just enough for galactic navigation.

Another feature to examine is the orientation of pulses. Do MSXPs cover the whole galaxy? It would make most sense to beam preferentially in the galactic plane, rather than in random directions. Indeed, ETIs are likely to live where stars and planets are, i.e. in the galactic disk, rather than in cold space. The astrobiological view thus predicts a preferential beaming orientation towards the galactic plane. This needs to be compared with the natural, astrophysical reasons why pulsars would beam in the galactic plane.

It is well known that about half of MSPs are found in globular clusters (Lyne and Graham-Smith 2012, 105). The high-density population of stars in globular clusters make stellar encounters likely, and therefore the formation of binary systems and eventually MSPs, which all support the standard astrophysical formation models. The astrobiological interpretation should thus look for different properties of MSPs, whether they are in globular clusters or elsewhere. It turns out that MSPs in the galactic disk have two subpulses that are 180° apart, while it is not the case in globular clusters (Chen, Ruderman, and Zhu 1998).

Another way to explore whether MSPs in globular clusters have anything to do with ETI would be to look at their beaming direction. Is it random? Does it overlap with other pulsar beaming? Is the beaming advantageous for galactic navigation?

Furthermore, the SETI-XNAV hypothesis may help to find new binary millisecond pulsars. The first step would be to model the coverage of MSPs, and search in places where the coverage is non-existent, not known. The prediction is that one should find a MSP filling this coverage gap.

What are the energy requirements for an effective PPS? This is a related issue that may lead to new lines of inquiry that we will now explore.

## 4.2 Power distribution

> *beacons are likely to be pulsed,*
> *both to lower the cost and*
> *to make the signal more noticeable*
>
> (J. Benford 2010)

From an astrobiological perspective, we would expect low-density regions of the galaxy to be populated by powerful MSPs, and regions where many MSPs are already present to be populated by less powerful ones. The reason is that the signal doesn't need to be strong in higher MSP density areas, such as globular clusters. MSPs would have different power ratings in order to allow an efficient coverage. Studying MSPs that are further away from the galactic plane, and checking if their beaming is indeed stronger is a way to test this idea. Existing studies of cost-optimized interstellar beacons may be useful for starting this project (J. Benford, Benford, and Benford 2010; G. Benford, Benford, and Benford 2010). Figure 6 shows a map of the distribution of MSPs. Is the spatial and power distribution random or particularly fit for galactic navigation?

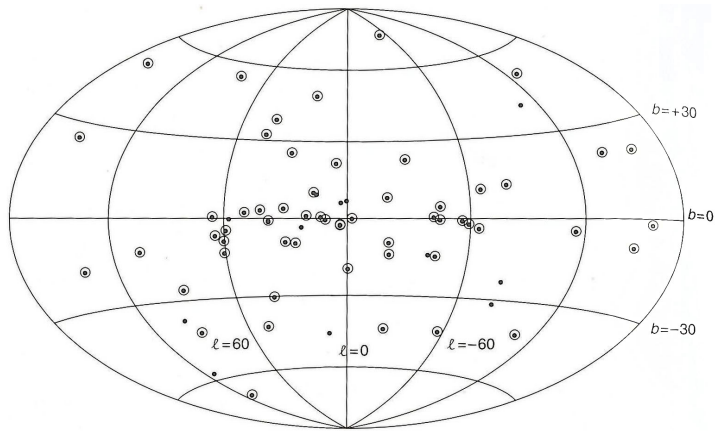

Figure 6 - The distribution of MSPs in Galactic coordinates, excluding those in globular clusters. Binary MSPs are shown by open circles. From (Lyne and Graham-Smith 2012, 116)

### 4.3   Population synthesis

On Earth, a GNSS such as GPS needs at least 24 satellites to provide Earth with full navigation coverage. How many pulsars would be needed to cover the galaxy? We would need to estimate the number $N_{pred}$ of MSPs that are needed to navigate the whole galaxy. $N_{pred}$ can then be compared with the actual number $N_{obs}$ of MSPs suitable for navigation. If the $N_{obs}$ of MSPs is close to $N_{pred}$, it tends to supports the ETI interpretation; and if not, it tends to falsify it.

The astrobiological prediction may also contrast with current astrophysical predictions of MSP population. MSP population is estimated to be between ~ 30 000 and ~ 200 000 (see respectively Lyne et al. 1998; Lyne and Graham-Smith 2012, 116).

There are a number of limitations and potential difficulties with this test. In terms of engineering and robustness, the theoretical number $N_{pred}$ would only provide a lower boundary. Indeed, even if there is just one PPS, it's likely that there will be more pulsars than the strict minimum. To illustrate this point, GPS has 30 satellites in orbit, whereas it only needs 24.

Another possibility is that there are different PPSs, like we have different GNSSs on Earth's orbit. This could lead to searching for different PPSs signatures, for example, looking at subsets of pulsars having typical common signatures, and that operate as a consistent PPS. In this case, we can still expect the total number $N_{obs}$ to be a *multiple* of the minimal number of pulsars to make a galactic navigation system. In the ideal case, we might deduce the number of PPSs, if there is a certain multiple of redundant MSPs (e.g. here on Earth, we could hypothesize that there are about 3 navigation systems in operation).

The situation could even be subtler if there are PPSs currently being set up, in the same way as the European Galileo GNSS is unfinished. In that case, the predictions would be harder to check, unless we have a method to distinguish between different PPSs, or to distinguish an operational versus a nonoperational PPS.

## 4.4   Evolutionary tracks

Evolutionary models that attempt to account for the diversity of accreting binary millisecond pulsars involve several non-trivial steps (see e.g. Lamb et al. 2009). In particular, the standard scenario for MSP evolution can not produce the X-ray MSP population (Kízíltan and Thorsett 2009; 2010). Another limitation of the standard scenario is the discovery of a MSP with an unexpected metal-rich companion (Kaplan et al. 2013). Generally, one hallmark of complex objects is that they have a rich causal history (Bennett 1988; Delahaye and Vidal 2016). Binary stellar evolution is clearly richer than single star evolution, and this may be an invitation to consider scenarios where ETI is involved.

I first became interested in pulsars in binary systems following the lead of the stellivore hypothesis, according to which some binary star systems in accretion might be considered as living (Vidal 2014, chap. 9; Vidal 2016a). The hypothesis starts by noticing that some accreting binary star systems (not only pulsars) display non-equilibrium thermodynamic features, such as energy budgeting, alternating ingestion and extrusion of materials. Such features can be interpreted as the hallmark of a metabolism, a common denominator of all life. Under this framework, "living" pulsars or pulsars under ETI influence are in binary systems, and isolated MSPs don't fit easily into the stellivore interpretation. Indeed, without an energy source (the companion star) to regulate the spin or magnetic field, the pulsar stability may be degraded with time. It is worth noting that the existence of single MSPs is also poorly understood by astrophysical models (Lorimer and Kramer 2005, 30).

From this astrobiological, stellivore hypothesis, one can attempt the following conjectures. First, isolated MSPs may have their stability degrade faster than binary MSPs. Second, binary MSPs may regulate their spin and magnetic field by using free energy from the companion star. A third expectation is that single MSPs would migrate toward an energy source – another single star. We may predict that they aim at the *nearest* single star to rejuvenate themselves and spin faster to join a PPS. The proper motion of single MSPs may thus be observed to be higher than binary MSPs. A fourth option is that single MSPs would be the equivalent of our space junk. Since they have no energy source, they are simply not working properly anymore. In that case, one might expect redundancy in single and binary MSPs coverage, or that the pulse characteristics of single MSPs would be different from the binary MSPs. Their suitability for navigation would thus be limited.

Current XNAV solutions use both single and binary millisecond pulsars. The stellivore hypothesis suggests that X-ray pulsars in binary systems will be shown to be more precise, reliable, robust, or important according to some navigation criteria, than other isolated MSPs.

The typically low magnetic field of MSPs may increase as a way to trigger accretion in a controlled manner. For example, MSP J1740-5340 could potentially swiftly switch from a radio phase to an accretion phase (Burderi, D'Antona, and Burgay 2002).

Some pulsars might evolve towards the millisecond range through ETI intervention. These pulsars would become more stable to contribute to a PPS. A prediction is that new stable pulsars would settle in regions covering new areas of the galaxy, or provide redundancy where there was previously no redundancy.

## 4.5 Synchronization

From the perspective of living systems theory (J. G. Miller 1978; J. L. Miller 1990), a *timer* is an essential subsystem in living things, providing key information for decision-making, to start, stop, change the rate, advance or delay processes, thus coordinating them in time.

A central characteristic of timers is their ability to reset. Biological clocks have this ability which is essential because the duration of day and night changes throughout the year. The world clocks are also regularly re-synchronized by time signals from radio stations around the world. In particular, the time on GNSSs satellites are synchronized by a *control segment* that communicates bi-directionally with the *satellite segment* (see Figure 7).

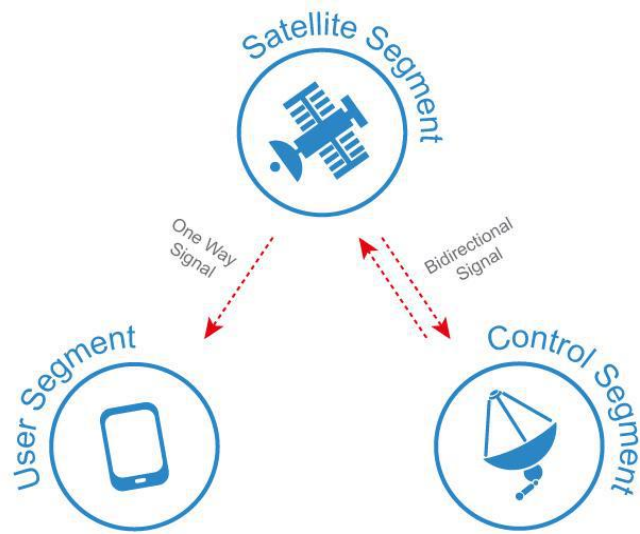

Figure 7 - Three segments of GPS. The *user segment* needs only a one-way signal to operate, while the *satellite segment* needs to communicate bidirectionally with a *control segment*. In the case of a PPS, the user segment would be the equivalent of a spacecraft, the satellite segment of a network of pulsars, and there are no known control segment. This analogy with GPS thus invites to look for a control segment in the galaxy. Figure from (Anver and Vasyl 2014).

Could we try to specify and observe a resetting process, or the equivalent of a control segment operating in a pulsar network? This is what I suggest to study here.

The fastest and most stable MSPs might constitute such a control segment, to which the other pulsars would synchronize. Concretely, we could look for time correction signals broadcasts (that exist in GNSSs), or synchronization waves. For example, synchronization might occur first on pulsars nearest the putative control segment, and then diffuse to further away pulsars. This could be investigated via rare MSP glitches, or other remarkable features, such as giant pulses in MSPs.

Humans have devised two major approaches for time synchronization, *centralized* or *distributed*. The centralized option is simply a central server dictating time. The distributed option is mostly used on the Internet and works with a *network time protocol* (Figure 8). Clock synchronization in a distributed system is far from obvious (Tanenbaum and Steen 2007). Given the size of the galaxy, a distributed solution is more likely than a centralized one. Various ways to synchronize should thus be explored to see if one of them is or could be taking place between MSPs.

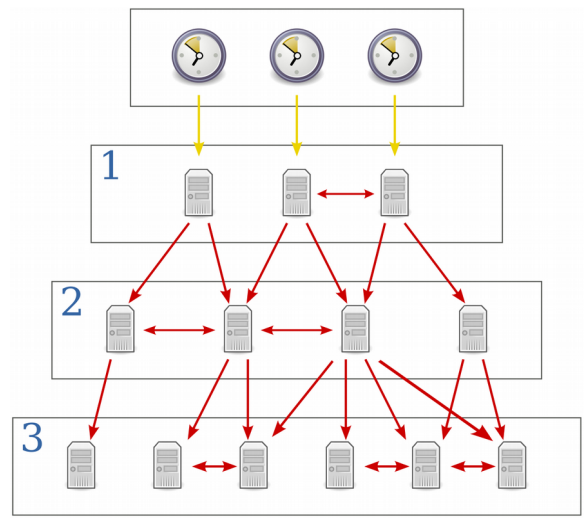

Figure 8 - Illustration of the network time protocol used to synchronize clocks. Illustration by Benjamin D. Esham.

## 4.6   Navigability near the speed of light

The issue of guiding fast objects with a GNSS is not obvious on Earth. A suitable navigation system for ETI should be able to provide effective guidance near the speed of light. In space, the distances are so immense, that any serious galactic traveler would likely travel at relativistic speeds. Does XNAV offer easy navigation near the speed of light? The astrobiological hypothesis predicts that it would.

## 4.7   Decoding galactic coordinates

A GNSS satellite broadcasts two critical kinds of information: the *time* of its local atomic clock, and its *position* relative to the center of the Earth. We already know that MSPs provide the equivalent of atomic clocks, but could it be that they also broadcast time-stamps and their position?

For short navigation missions inside the solar-system, such coordinate broadcasting wouldn't be needed (Buist et al. 2011, 155b). However, for long trips, occasional pulsar ephemeris updates would be suitable (Graven et al. 2007). Indeed, even if the MSP population has a lower velocity than the normal pulsars (see Table 1), they still move in the galaxy, and this needs to be taken into account.

The astrobiological consequence is straightforward. We could seek ephemerides updates, broadcasted by MSPs themselves! The question of "how?" requires additional working assumptions. First, we can reasonably expect that the logical reference frame for galactic navigation would be the galactic center. We could look for traces of coordinates in a variety of ways: polar, cartesian, vectorial, etc.

Machine learning algorithms could be used to try to find ephemeris data in pulsars. For example, an algorithm could be trained by taking the average pulse profile or polarization data as input, and output the position. The validity of the method could be tested with raw data from GPS or GLONASS satellites. For example, it should be able to deal with various coding schemes such as frequency division multiple access (FDMA, initially used by GLONASS) or code division multiple access (CDMA, used by other GNSSs).

If the picture I sketched in this paper is correct, coordinates should only be broadcast by MSPs, not by normal pulsars. This provides a null hypothesis.

Another way to test the ephemeris broadcasting idea is to focus on moving MSPs, and test if there is a correlation between what they broadcast and their galactic coordinates. The astrobiological prediction here is that the average pulse profile or other broadcast data would change depending on position.

## 4.8 Directed panspermia: navigation and propulsion

Directed panspermia is the suggestion that advanced extraterrestrial beings have the ability and motivation to seed the universe with life (Crick and Orgel 1973; Crick 1981). Broadly speaking, the *generalized principle of cosmic reversibility* (inspired by Crick and Orgel 1973), says that:

If we are capable of doing X in the future, then, given that the time was available, another extraterrestrial civilization might well have done X already.

If we take X to be "life seeding", we humans can already think about many reasons to seed the universe with life (see e.g. Mautner 2004). Following this principle, ETI might have done it already. For such life-seeding purposes, Mautner (1995) identified the necessity of extremely precise astrometry, toward a resolution of 0.1 milliarc-seconds. However, thanks to XNAV, there is now a straightforward navigation solution.

The present Breakthrough Starshot project[1], attempting to reach the nearest star, Alpha Centauri, raises another issue beyond navigation, namely the *propulsion* of probes. The current solution consists in using powerful lasers installed on Earth to beam the necessary energy to the probes. Applying the principle of cosmic reversibility, if ETI is beaming propulsion power, this may be observable (J. Benford and Benford 2016). However, as

1 https://breakthroughinitiatives.org/Initiative/3

the probes are further and further away, it is harder and harder to accurately beam toward them.

A propulsion solution might be *to use pulsars not only for navigation, but also for propulsion (Vidal 2016b)*. Navigating the galaxy requires a lot of fuel, and it has often been proposed that an external power source could be used (e.g. Crick 1981, 133; Guillochon and Loeb 2015; Dyson 2015 at 23:35). The advantages of delegating both navigation and propulsion to pulsars would be immense, as it would make navigation fully autonomous. No communication with the point of departure is needed, and the spacecraft doesn't need to carry its own fuel, so it can remain very light and small. The beaming of pulsar energy might thus be able to make RNA, DNA strands, viruses, microbes, bacteria, other microorganisms or probes travel through space. The requirements are that the entities should be able to survive in space, be propelled with pulsar radiation power, and navigate with a PPS. The feasibility remains to be examined. But even if the propulsion is weak, and the entities travel slowly, provided the seeding project is not in a hurry, it may work.

A core advantage of GNSSs is that the navigation system is *passive*, and requires few or cheap equipment on the user segment side: no atomic clock or sophisticated antennas are needed on the user segment (Spilker and Parkinson 1996, 30). The same passive quality holds for PPS. Adding fuel passivity would lead to the equivalent of self-driving cars powered by solar panels.

A probe or seed using PPS could go anywhere in the galaxy, with an accuracy of 100 m! Actually, the idea to use radiative pressure as a way to propel dates back to the birth of the scientific formulation of panspermia (Arrhenius 1908). The original idea here is to use pulsars both as a propulsion *and* a navigation solution.

As weird as it may seem, we are already able to test a range of such directed panspermia scenarios by testing if some particularly space- and acceleration-resistant organisms (Wikipedia contributors 2017) could use or detect beams mimicking MSXP radiation. Again, the null hypothesis would be to test normal pulsar radiation, in which case we would expect no reaction. This may be achieved today with an experimental setup using the Goddard X-ray Navigation Laboratory Testbed, also known as "the pulsar-on-a-table". A similar experiment has already been performed, using a planetarium to show that beetles can orient themselves thanks to stars (Dacke et al. 2013).

Another test consists in checking if galactic coordinates of our solar system would be encoded in known space-proof microorganisms. Of course, we should take into account that our solar system is revolving around the center of the galaxy. In the best case, this could even indicate *when* the organism arrived on Earth, by computing the orbit of our solar system around the galaxy, and comparing the coordinate found in the entity with our current galactic position.

It might also be possible to distinguish between level 1 and 3 in the artificiality of pulsar navigation (see Table 3). The prediction is that if we find pulsar ephemerides, plus the destination, then the PPS is likely to be natural (level 1). However, if we find the destination only, it may mean that no local ephemerides data was needed, and that pulsars do broadcast ephemerides, and therefore that they are artificial (at least level 3).

Although not very convincingly, some authors have speculated that there might be a message in the DNA of organisms (Yokoo and Oshima 1979; Nakamura 1986), or in the genetic code itself (shCherbak and Makukov 2013). Recent progress in DNA computing and DNA data storage suggests that even if DNA is slow to write and read, it has other qualities such as long data retention, and a data density carrying capacity orders of magnitudes higher than today's best storage technologies (Extance 2016; Erlich and Zielinski 2017). The proof of galactic-scale ETI-engineering might lie right inside known DNA!

## 4.9   Information content

*Has anyone examined systematically*
*the sequencing of pulsar amplitude and polarization nulls?*
Carl Sagan (in Dyson et al. 1973, 228)

*The most rational ET signal would be a series of pulses*
*that would be evidence of intelligent design*
Frank Drake (in Swift 1990, 64)

In a discussion about the possibility of astroengineering, Carl Sagan proposed to analyze pulsars' pulses for messages or traces of artificiality. This reminds us that SETI through pulsars need not be limited to navigation.

As Drake notes, pulsed radiation would technically make most sense for purposes of communication. This also holds economically, as we mentioned earlier (see section 4.2).

Communication implicitly assumes a shared context, a shared reference system. We would not try to decipher dolphin communication by just looking at the recorded patterns they emit. Instead, what ethologists typically do is study communication in the broader context of their natural habitat, to try to see correlations between communication and typical behaviors such as eating, mating, competing, playing or warning. If we want to have a chance to decipher putative ETI signals, we need to know the shared environment. *That environment is clearly our Milky Way Galaxy*. So we must think of SETI in this astronomical and astrophysical context (see also section 5 and Cordes and Sullivan 1995).

In terms of information patterns, regular radio signals on Earth come from navigation beacons, time standards identification, transponders and radars (Wolfram 2002, 1188b). Wolfram notes that such signals "sound uncannily like pulsars"! Irregular signals are harder to decode, as they can be modulated in many different ways, and require cryptanalysis. But irregular signals can also reflect a high bandwidth of information transmission.

A general counterargument to the possibility of SETI is that if ETI communicate effectively, they would send compressed information, which would look random. Indeed, as Wolfram (2002, 836) writes, "regularity represents in a sense a redundancy or inefficiency that can be removed by the sender and receiver both using appropriate data compression".

The situation is different with a navigation system. Navigation signals have a different purpose than a private communication. A navigation system has fair chances to be without encryption because it is likely to be for everybody, and therefore anti-cryptographic (see e.g. Dixon 1973). Another reason why anti-cryptography would be suitable is that it would allow a low computational and energetic load on the user segment side.

The SETI-XNAV program is a promising area of focus for SETI, as navigation signals are more regular and easier to process than other potentially irregular, highly modulated or encrypted ETI signals.

The beginning and end of communication would likely have markers, like any file in a file system, even if they are encrypted. This is why even assuming pulsars are used for communication, it makes most sense to try to decipher a timing and navigation system first. In a way, we already did discover how to use pulsars for navigation in 1974!

Because of the first lighthouse model, pulsars are often said to be galactic lighthouses. On Earth, lighthouses obviously emit visible light. But this represents only the tip of the informational iceberg, as in a modern harbor, much more information is communicated at other invisible wavelengths: radio communication occurs between boats, or between boats and the harbor. In the same way, pulsar navigation may be the obvious part, while other information-rich communication may be going on.

The challenge to decode other possible messages in pulsars is thus open. Since pulsars are distributed, it makes sense to look for concepts and tools coming from distributed computer science. For example, for space missions, a network architecture has been proposed, *delay-tolerant networking*, that can accommodate heterogeneous connections, as well as discontinuities in network connectivity (see e.g. Burleigh et al. 2003). A decentralized mesh network might also make sense for pulsar communication.

Recently, it has been shown that multiplexing could use an additional degree of freedom, by using orbital angular momentum (Wang et al. 2012). The results are impressive and promise future wireless communication reaching speeds up to 2.5 terabit per second! Since pulsars are rotating neutron stars, if they are modified or used by ETI, it might be that such kind of highly efficient multiplexing is used.

Modulating circular polarization is a method used in X-ray communication. It might also be used in MSXPs. Today we have the capabilities to study individual pulses of MSPs, their varying intensity or polarization modes (e.g. Osłowski et al. 2014; K. Liu et al. 2016). There is thus room to study MSPs with all the SETI methodology. For example, linear feedback shift register is used in GPS to generate ranging codes, and might thus be also used by ETI (Wolfram 2002, 1190). Another potential test is the Kullback–Leibler divergence measure, that has already been used in astrobiology (Wright et al. 2016) and could be applied to pulsars. Any other method to test pulsars' signal complexity would be a valuable attempt (for a review on SETI signal detection and analysis, see Siemion et al. 2015).

# 5   Pulsars as standards

In their landmark SETI article Cocconi and Morrison (1959) argued that the radio emission line between 1.420 GHz and 1.666 GHz is a unique *frequency window* for all civilizations in the galaxy. In this part of the radio spectrum there is little noise from natural celestial sources, so artificial transmissions would easily be distinguished by any civilization that may have emerged in the galaxy.

Cordes and Sullivan (Cordes and Sullivan 1995; Sullivan and Cordes 1995) introduced the concept of *astrophysical coding*, arguing that intentional signals are likely to be encoded and detectable with the same techniques used by regular astrophysics. Astrophysical coding thus supposes a shared astronomical and astrophysical methodology amongst civilizations in the galaxy. In particular, pulsars would already be known and observed by other ETIs, so searching for an artificial signal related to pulsar's properties is a promising search strategy (see also Dixon 1973).

The usefulness of pulsars makes them likely to be used as standards by all putative civilizations in the galaxy. In contrast to the SETI-XNAV quest (section 4), this section remains valid starting from Level 1 of pulsar artificiality (see Table 3), i.e. that all pulsars are natural, but may be still used as standards by ETI. In all likelihood, advanced civilizations in our galaxy would have discovered the timing, positioning and communication uses of MSPs.

## 5.1   Frequency window standard

Heidmann and collaborators (Heidmann 1988; 1992; Heidmann, Biraud, and Tarter 1992) proposed to convert the rotational frequencies of pulsars into radio frequencies in a well-defined and quasi-unique way. This is a promising start for SETI because by using pulsars as a shared, galactic standard, it reduces search space in terms of bandwidth. However, it can be criticized because it also assumes using a mathematical constant to convert the rotation frequency into a bandwidth, a step which seems somehow arbitrary.

## 5.2   Pulse window standard

Which timescale should we use if we want to communicate or listen in the galaxy? There are many possible timescales, but Sullivan (1991) has proposed, in addition to the preferred *frequency window*, a preferred *pulse window*, based on observed pulsar periods. He suggested that they would constitute a natural galactic communication channel.

## 5.3   Pulsars and habitable stars alignment

In a similar line of thinking, Edmondson and Stevens (Edmondson and Stevens 2003; Edmondson 2010) proposed a SETI-strategy based on pulsars, where alignments are sought between at least planet Earth, a pulsar, and a habitable star. A habitable star is defined as one that could support life on one of its planets. Thanks to this special alignment, the pulsar is used as a beacon to attract attention for us to further look for a signal coming from a planet of the aligned habitable star in question.

Remarkably, Edmondson found up to 6 pulsars aligned with 6 habitable stars (see Figure 9). Intriguingly, all 6 pulsars are MSPs, and in binary systems (Lynch et al. 2012), thus appearing in the stellivore configuration.

M62
O- - - - - - - - - - - ->o<- - ->o<- - - ->O<- - ->o<- - - ->o<- - - ->o
J1701-3006A-F             83522    83241    Earth    23302    23291    23438

Distances: HIP83522 @ 88pc;  HIP83241 @ 79pc;  HIP23302 @ 41pc;  HIP23291 @ 122pc;  HIP23438 @ 219pc

Figure 9 - Six pulsars in globular cluster M62 are aligned with 6 habitable stars, of which Earth is one. Distances are given for the locations of the stars relative to Earth.
(From Edmondson 2010)

However, as recognized by the authors, this proposal is asymmetric, in the sense that the ETI transmitting invests much more than us, just to establish contact. What motivation could an ETI have to ensure that we (in particular!) discover their existence? It may be anthropocentric to think that we are worth so much effort and attention. The same applies to any civilization that has not discovered ETI yet. From an advanced ETI perspective, what is the worth of helping other civilizations discover that they are not alone?

Our arguments about the ease of directed panspermia thanks to pulsar navigation (see section 4.8) suggest another interpretation. Such aligned pulsars would beam not information, but energy to propel life-seeding probes towards habitable worlds, in order to pursue the most universal enterprise: build complexity and reduce the entropy of the universe (for more arguments about this ethical imperative, see e.g. Mautner 2009; Vidal 2014, 271–74).

## 5.4   Timing and positioning standard

*The clock, not the steam engine,*
*is the key-machine of the modern industrial age.*
(Mumford 1934, 14)

When Galileo started his scientific experiments, he was able to keep track of time, and such precise timekeeping was a key ingredient to the birth of modern science. In society, as Mumford (1934) analyzed, clocks were central to the industrial revolution, because "the clock is not merely a means of keeping track of the hours, but of synchronizing the actions of men" (Mumford 1934, 14).

We already mentioned that living systems at all scales require a *timer* subsystem to coordinate their actions (J. L. Miller 1990). For galactic timing purposes, *any form of life is likely to use pulsars as a timer subsystem*, whether pulsars are natural or artificial.

But even here on Earth, the case to develop a pulsar-based time-scale is strong (Hobbs et al. 2012). A pulsar time-scale presents at least the following advantages (Hobbs 2012, 2781a):

> (i) an independent check on terrestrial time-scales using a system that is not terrestrial in origin,
> (ii) a time-scale based on macroscopic objects of stellar mass instead of being based on atomic clocks that are based on quantum processes and
> (iii) a time-scale that is continuous and will remain valid far longer than any clock we can construct.

Although in the 1990s atomic clocks and pulsars were comparable in stability (e.g. Rawley et al. 1986), this is not the case anymore. Nowadays, optical clocks have surpassed both atomic clocks and MSPs in stability (see Figure 10). However, as shown in the trend-line of Figure 10, long observations of the stablest MSPs may lead to comparable or better clock stability. Indeed, the observed pulsar slowdown rate (increase in $\dot{P}$) may be due to a relativistic effect called the Shklovsky effect (Shklovsky 1970). The observed pulsar increase in $\dot{P}$ would not be intrinsic, but related to the pulsar's distance, and its proper motion (for more details, see Camilo, Thorsett, and Kulkarni 1994; Lyne and Graham-Smith 2012, 52). By contrast, even optical clocks are limited in stability, represented by the solid arrow on the bottom-right of Figure 10.

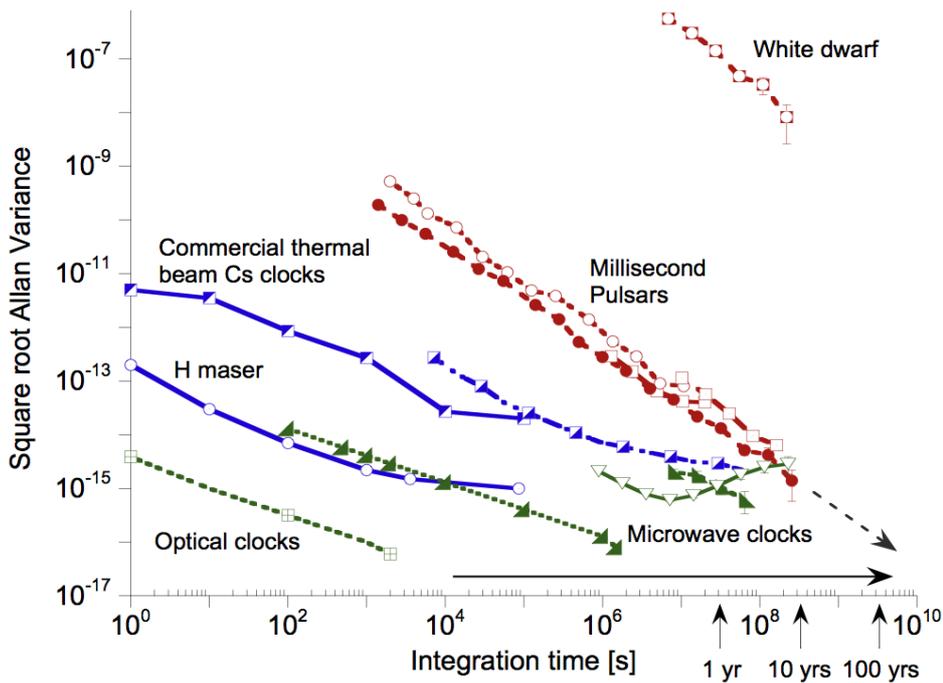

Figure 10 - Comparison of terrestrial and astrophysical clocks. The frequency stability is expressed in terms of square root Allan variance (y-axis), and the x-axis represents the integration time. Optical clocks are currently the best ones, but the long-term trend of the best millisecond pulsar (dashed arrow) shows that they may compete if we allow a long integration time. See (Hartnett and Luiten 2011) for details.

Even granting that Earth clocks will remain more stable than astrophysical clocks, any Earth clock is subject to catastrophic risks. By contrast, pulsars are stable through millions or hundreds of millions of years. This stability in time suggests that we could benefit from establishing a pulsar timing standard starting from today. Actually, two pulsar clocks have already been installed in 2011, in St Catherine's Church, Gdańsk, Poland, and in the European Parliament in Brussels, Belgium (Gdańsk Tourism Organization 2017). The very existence of the international pulsar timing array is also a promising step in this direction.

MSPs may thus constitute a galactic timing and navigation standard for all civilizations in the galaxy. We already mentioned some lines of inquiry that can be tested in the context of directed panspermia. Yet, as with GNSSs, timing standards have many more applications beyond navigation. A similarly wide array of applications might hold at a galactic scale, and remain to be explored, even if pulsars are perfectly natural (Level 1 in table 3).

## 5.5 Communication standard

Pulsars could also be key in providing metadata standards for any communication. On Earth, any letter or email contains metadata information about where it comes from, where it goes, and when it was written. We can reasonably suspect that similar conventions exist regarding any potential galactic communication. Most messages are likely to be galacto-tagged, and pulsar-time-stamped, and they are likely to be so by reference to MSPs. This simple remark is constraining for SETI. Given any suspicious message we want to decode, the first step becomes to attempt to decode not the message itself, but its timing and galactic origin and destination metadata.

If we want to engage in messaging to extraterrestrial intelligence (METI), a pulsar communication standard should be taken into account. If we want to, we are able to easily locate ourselves in space-time. If ETI made a PPS, what are the policy issues? Do we have the right to use it without asking permission (for a discussion, see Vidal 2017, sec. 7.7)? I am just scratching the surface here, but the consequences of pulsars as standards are fundamental and far-reaching, both for SETI and METI.

# 6 Discussion

At this point, the reader may have many objections. Let us address a few of them.

## 6.1 SETI-XNAV, a modern bias?

Isn't this SETI-XNAV research agenda just a bias of seeking our modern technology in the sky? In the late $19^{th}$ century, canals were revolutionizing water and sewage distribution management, and scientists were seeing canals on Mars (Dick 1996). In the 1960s, radio and TV were a revolution, so traditional SETI started to look for radio signals in the galaxy. Today, we have GNSSs, and we start to think of PPS as an engineered navigation system.

This objection is quite valid, and the bias is real. We have always looked for alien technologies taking as a guide the state-of-the-art technology of a certain epoch. However biased, this heuristic makes sense, as it probes ideas related to our latest operational technological breakthroughs, rather than being based on old or imaginary ones. It would seem outdated to look for canals on exoplanets, and it would seem too speculative to try to find ways to detect a wormhole structure throughout the universe.

A strong argument is that navigation is a universal need, namely the need, *for whatever reason*, to go from one place to another. We emphasize again that a timer is a fundamental living system component, and if there is life in the galaxy, the timer is likely to be used to navigate, coordinate actions and communication. There is not much speculation involved in postulating that navigation and timing needs are universal.

Finally, we did not find artificial canals on Mars, we did not find traces of intelligent radio signals, but we did find out how to use pulsars for navigation.

## 6.2 How (im)plausible is pulsar engineering?

The idea of extraterrestrial engineering at a stellar or galactic scale may seem unlikely if not impossible. However, this belief ignores our broader galactic and cosmological context.

Back in 1975, Hart (1975) showed that even at 10% of the speed of light, the galaxy could be traversed in 650 000 years. Relative to the age of the galaxy, other spreading models support the possibility of a quick intelligent life expansion (e.g. Newman and Sagan 1981; Armstrong and Sandberg 2013).

To this we can add many independent arguments showing that if ETI exists, it is likely much older than us (see e.g. Crick 1981, chap. 10; Lineweaver 2001; Lineweaver, Fenner, and Gibson 2004; Dick 2009, 467–68; Loeb 2014). The difference between ETI and us could be *billions of years*, i.e. a difference in development comparable to the time needed to evolve from unicellular organisms to complex technological society.

An important objection to galaxy-wide engineering is that outer space growth is uninteresting to conquer, since it is mostly empty, long and expensive to travel through. The return on investment would be too low. Instead, intelligent life would develop into inner space, taking control of smaller and smaller scales, higher densities, and higher energies (Barrow 1998, 133–38; Smart 2009; 2012; Ćirković 2008; Vidal 2014, chap. 9; Last 2017).

However, inner and outer growth are not mutually exclusive. This is illustrated with the internet or GNSSs, which rely on microscopic technologies (computers, atomic clocks), and yet accelerate development and complexification on a planetary scale.

Even for a civilization developing inwardly, there are at least two reasons why space navigation might be of interest, if not of necessity. First, for any Type II civilization on Kardashev's (1964) scale, using stellar energy or collecting other galactic resources would require one to move. Even if a civilization is not interested in outer space, it may *have to* travel through space to get resources for its inner growth.

Second, a life-affirming ethics may encourage advanced civilizations to spread life, by seeding life to other solar systems, instead of leaving them lifeless. This could be achieved relatively cheaply through the sending of small probes or microorganisms.

These are just basic reasons, but there may be many more, such as galactic mapping or military reasons. Indeed, if space travel becomes common, it may in turn motivate setting up on a galactic scale what we have done with GNSSs on an Earth-scale.

## 6.3    Who could have done it?

One may object that one extraterrestrial civilization wouldn't have had the time to set up such a huge navigation system. The galaxy is so big that it would be a too huge investment to be feasible.

However, this is an anthropocentric objection. There is no reason that the lifetime of ETI should be in the same order of magnitude as a human life (100 years), or even a human civilization (1000 years). An advanced, surviving civilization might live millions or hundreds of millions of years. It might also continue the work of prior civilizations.

To set up an artificial PPS, there are two possibilities: either one single galactic civilization did it, or several did it. Even one single civilization setting up a PPS is plausible. Let us see why. Earth-like planets on other solar-type stars are on average 1.8±0.9 billion years older than the Earth (Lineweaver 2001). Bradbury *et al* (2011, 161) added that since 1.8 Gyr is an average, the galaxy is likely to be dominated by even older civilizations. Given that the galaxy is 160 000 lights years wide (Xu et al. 2015), in 2 billion years, it's possible for signals to do 6 250 return trips.

However, there is no need and no reason for one civilization to do all the work. Modern engineering tends to develop stigmergic, distributed, self-organizing and scalable architectures. It would be much more efficient and fair to let different civilizations add nodes in a PPS, a method that made the architecture of the Internet a success. We could imagine a distributed construction design, where information through electromagnetic waves would communicate the need and steps to set up additional pulsar beacons. To take Earth as an example, we could imagine receiving information about how to add a node in the PPS network by modifying the nearest pulsar.

The maximum age of ETI and the galactic spreading models thus leave a lot of space-time for intelligence to develop, and we must adapt our mindsets and SETI search strategies accordingly.

## 6.4   How could pulsar engineering work?

> *One would need only a very small movable shield*
> *above a pulsar surface to modulate emission to Earth.*
> *This seems much easier than generating*
> *an entire pulsar for communications.*
> *For signaling at night it is easier*
> *to wave a blanket in front of an existing fire*
> *than to start and douse a set of fires*
> *in a pattern which communicates a desired message.*
>
> Carl Sagan in (Dyson et al. 1973, 228)

Of course, we don't know how pulsar engineering could work. But some authors, including Carl Sagan, have speculated about this issue. On Earth, the use of smoke signals is effective for long-distance communication and has been used since ancient times. Sagan suggested that a similar method could be employed with pulsar radiation. More recently, Paul Davies (2010, 105) suggested that pulsar modulation could be achieved by a technologically savvy civilization who might…

> try using the pulsar emission itself to convey the message, by modulating the natural pulses in some way. That would neatly solve the power problem - pulsars are so powerful they can be detected across the entire galaxy with a modest radio telescope. The signal would then show up as a pattern in the frequency, intensity or polarization of the radio pulses.

Intelligently modulated signals would indeed save a lot of energy compared to setting up a powerful communication device from scratch. Interestingly, there is a model (Lyne and Manchester 1988) to explain the pulse components that supposes a patchy beam structure (Figure 11).

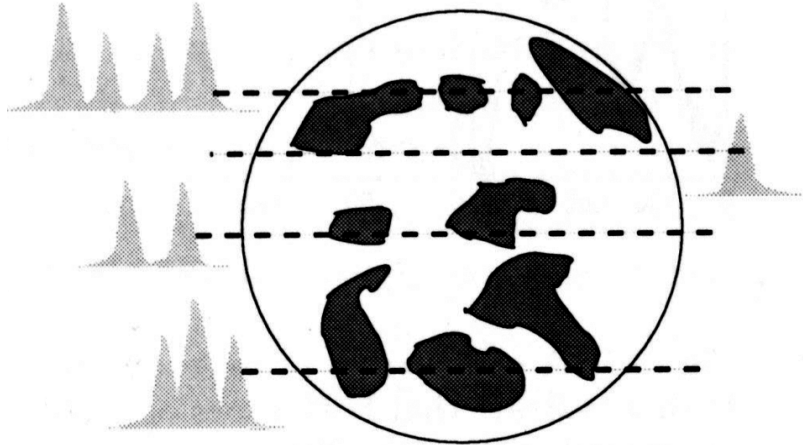

Figure 11 - A model to explain the pulsar's components suggests that the interior of the beam is filled with discrete emitting regions, that make only parts of the emission active. Figure adapted from (Lorimer and Kramer 2005, 73)

From the stellivore hypothesis framework, an energy source to modify the spin, magnetic field or emission mode of a pulsar would be needed, ensuring its proper function and maintenance. A companion star would provide such a necessary energy source. Pulsar engineering would be most active in binary systems in accretion.

Other ways an X-ray binary could have been modified by an advanced civilization have been proposed (Fabian 1977; Corbet 1997; Carstairs 2002). Recently, a remarkable model on how a pulsar's signal may be modulated by an artificial satellite has been proposed (Chennamangalam et al. 2015). The authors propose that studying the phenomena of nulling, mode-changing, pulse-to-pulse, intra-pulse variation, or sub-pulse drifting may show signs of non-natural processes. They predict that an artificially modulated pulsar would "exhibit an excess of thermal emission peaking in the ultraviolet during its null phases, revealing the existence of a modulator".

## 6.5   Is SETI-XNAV new?

As we already mentioned, the idea of making variable stars for navigation is not original, since it was explored in science fiction even before the discovery of pulsars (Smith 1957). Seth Shostak (1999) also speculated in a fiction essay that the discovery of ETI would be a time signal, similar to the WWV call sign of the United States' National Institute of Standards and Technology. Joseph Voros (2014) hypothesized that Hoag's object, a galaxy with a strange ring structure, might be due to extraterrestrial intelligence engineering. As a way to empirically test the idea, he proposed to look for time-keeping signal beacons that would act as a standard clock to synchronize activities. He further suggested that the signal would likely broadcast isotropically, and be directed along the galactic plane.

LaViolette (1999; 2006; 2016) can be acknowledged for having speculated that pulsars may be artificial, and might also constitute a navigation system. Unfortunately, these claims are combined with mythology, commit fallacious probabilistic arguments, do not lead to specific predictions, and are uninformed by modern XNAV research (see also Vidal 2014, 250 for more critique).

By contrast, the proposal of this paper is to *join pulsar astrophysics, astrobiology and navigation science to form a SETI-XNAV research program*.

## 6.6   What could the benefits be?

Even if SETI-XNAV fails, there are benefits to reap for the future of humanity. First, testing the hypotheses of the SETI-XNAV quest (section 4) will improve our knowledge of pulsars and XNAV. Second, our Earth-based space navigation method, such as the Deep Space Network may be augmented with XNAV (e.g. Graven et al. 2007). By contrast to human-made navigation solutions, pulsars are immune to solar flares, to humanly hostile attempts at disabling them, for they are very difficult to jam (Buist et al. 2011, 153b).

If XNAV could be made even more accurate *and* if we could relay such X-ray signals to Earth (e.g. by setting up a GNSS constellation translating X-rays to radio waves) then we could send cheap and small satellites doing this translation, extending the PPS for Earth. It would constitute a secondary navigation system, if our own would fail; or an augmentation system; or just a very cheap, robust and long-lived GNSS, even if it is not as accurate as current GNSSs.

Current GNSSs define position in a classical, Euclidian way, to which relativistic corrections are added. Instead, XNAV is an invitation to naturally use a Minkowskian positioning system instead of the current approach (see e.g. Tartaglia 2010; Coll, Ferrando, and Morales-Lladosa 2010; Bunandar, Caveny, and Matzner 2011). Applying such a true relativistic framework to GNSSs promises a cleaner mathematical treatment, and more ease to deal with relativistic effects.

Additionally, if PPS is natural, and works nearly as well as GNSSs, then it begs the question: why would our GNSSs need a control segment at all? Distributed, independent clocks might do well for navigation, and thus savings might be made on control segments of GNSSs.

Scientifically, SETI-XNAV is a concrete ETI hypothesis to test. The data is here, the timing and navigation functionalities are here. Historically, the suspicion of artificial canals on Mars triggered space missions to Mars and developed knowledge about Mars. Similarly, the project to try to decipher any potentially meaningful information in pulsar's signals (see section 4.9) could lead to the development of tools and methods that can be used for any future candidate signal.

# 7   Conclusion

The need of autonomous space navigation is timely, as our epoch is experiencing a renewal of space exploration, with numerous scientific missions inside the solar system, but also bolder missions funded by billionaires, such as to colonize Mars (Elon Musk), or to reach Alpha Centauri (Yuri Milner). To guarantee the success of such missions, accurate navigation will be critical, and pulsars are currently the best option to navigate the solar system and the galaxy with high accuracy.

Even if SETI-XNAV fails, the program still promises various benefits because it will contribute to XNAV research, and may help to augment or design more efficient and cheaper global navigation satellite systems here on Earth.

A broader outlook on convergent evolution suggests that GNSSs and PPS may one day be considered as instances of "cosmic convergent evolution" (Flores Martinez 2014). The convergence could come in two different forms. First, the time of arrival navigation method that humans developed for GNSSs is also applicable on a galactic scale thanks to pulsars, and other ETIs are likely to also converge on this navigation solution. Second, if the PPS was indeed engineered by ETI, it would be a remarkable case of technological convergence, on radically different scales: GNSSs provide an Earth navigation solution and millisecond pulsars a galactic navigation solution.

Conceptually, SETI-XNAV extends SETI to search for a *distributed* signal (an intelligently set-up navigation system), instead of searching for a *localized* signal around one particular star or planet. If the program succeeds, it would lead to the discovery of extraterrestrial intelligence, through their engineered navigation system. The discovery would cause a much anticipated scientific worldview shift, with a complex societal impact (see e.g. Dick 2015; Vidal 2015).

To sum up, this paper draws two major conclusions, one to be expected, the other still uncertain. First, all pulsars could be perfectly natural, but we can reasonably expect that civilizations in the galaxy will use them as standards (section 5). By studying and using XNAV, we are also getting ready to receive and send messages to extraterrestrial intelligence in a galactically meaningful way. From now on, we might be able to decipher a first level of timing and positioning metadata in any galactic communication.

Secondly, what remains uncertain is whether the pulsar positioning system is natural or artificial. We put forward the SETI-XNAV quest to answer this issue. It draws on pulsar astronomy, as well as navigation and positioning science to make SETI predictions. This concrete project is grounded in a universal problem and need: navigation. Decades of pulsar empirical data is available, and I have proposed 9 lines of inquiry to begin the endeavor (section 4). These include predictions regarding the spatial and power distribution of pulsars in the galaxy; their population; their evolutionary tracks; possible synchronization between pulsars; testing the navigability near the speed of light; decoding galactic coordinates; testing various directed panspermia hypotheses; as well as decoding metadata or more information in pulsar's pulses.

To emphasize the significance of the SETI-XNAV research program, imagine that we would find around an exoplanet's orbit well-distributed timekeeping devices with an accuracy comparable to atomic clocks, beaming timing information that can be used as a positioning system, just like GPS. Wouldn't we be compelled to explore the hypothesis that extraterrestrial intelligence is at play? This is exactly the current situation with millisecond pulsars, but on a galactic scale.

# 8   Acknowledgements

I thank Thomas Provoost for the suggestion to look for pulsar synchronization waves, Werner Becker, Jason T. Wright and two anonymous reviewers for corrections and constructive feedback, as well as Edward Cooke and Cadell Last for their careful proofreading.

# 10  Argumentative maps

The following argumentative maps present the core argumentation of this paper. The first graph maps the problem described in introduction and section 3, while the second graph maps the core argument presented in the paper. Please read in a top-down direction.

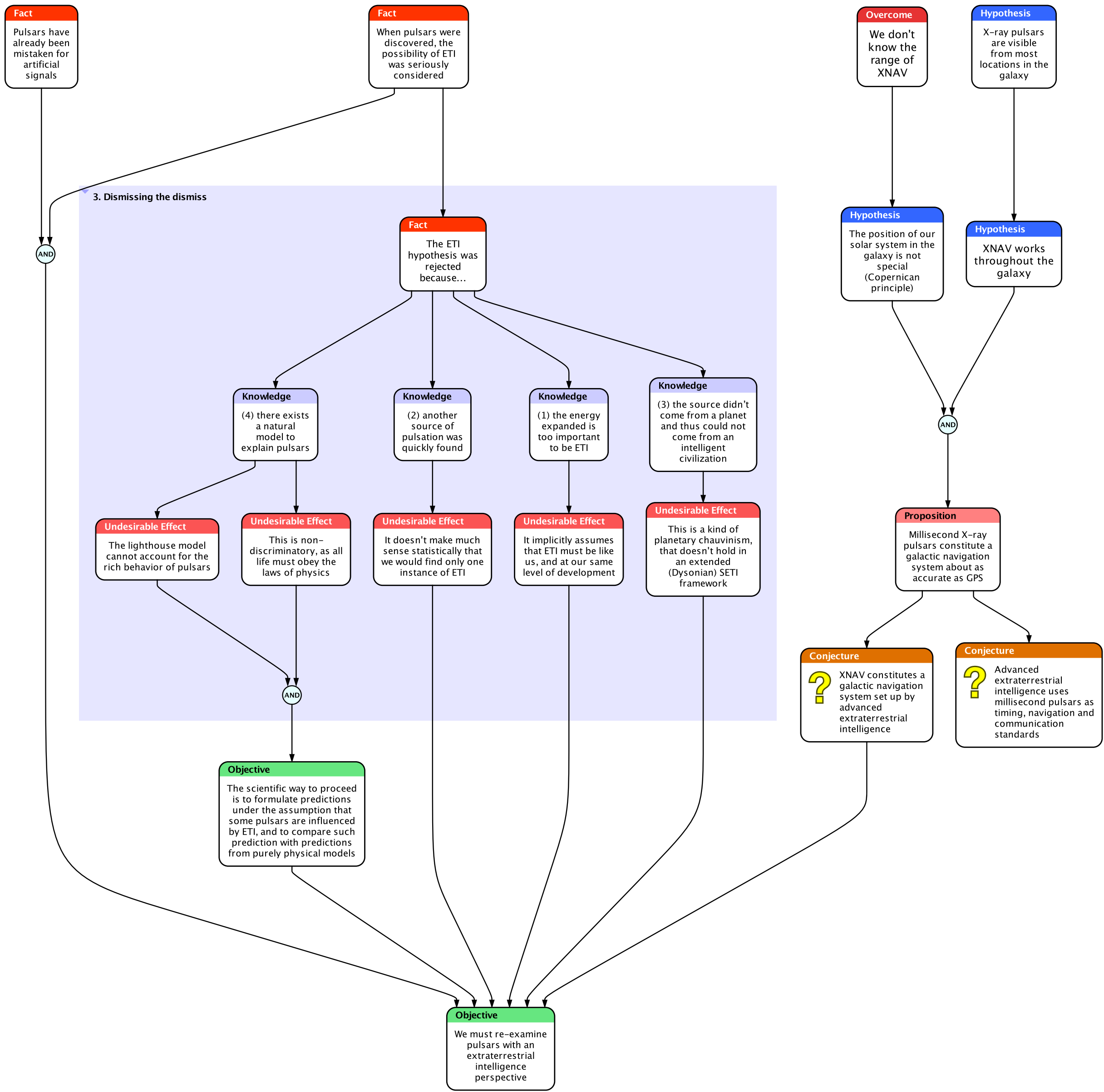

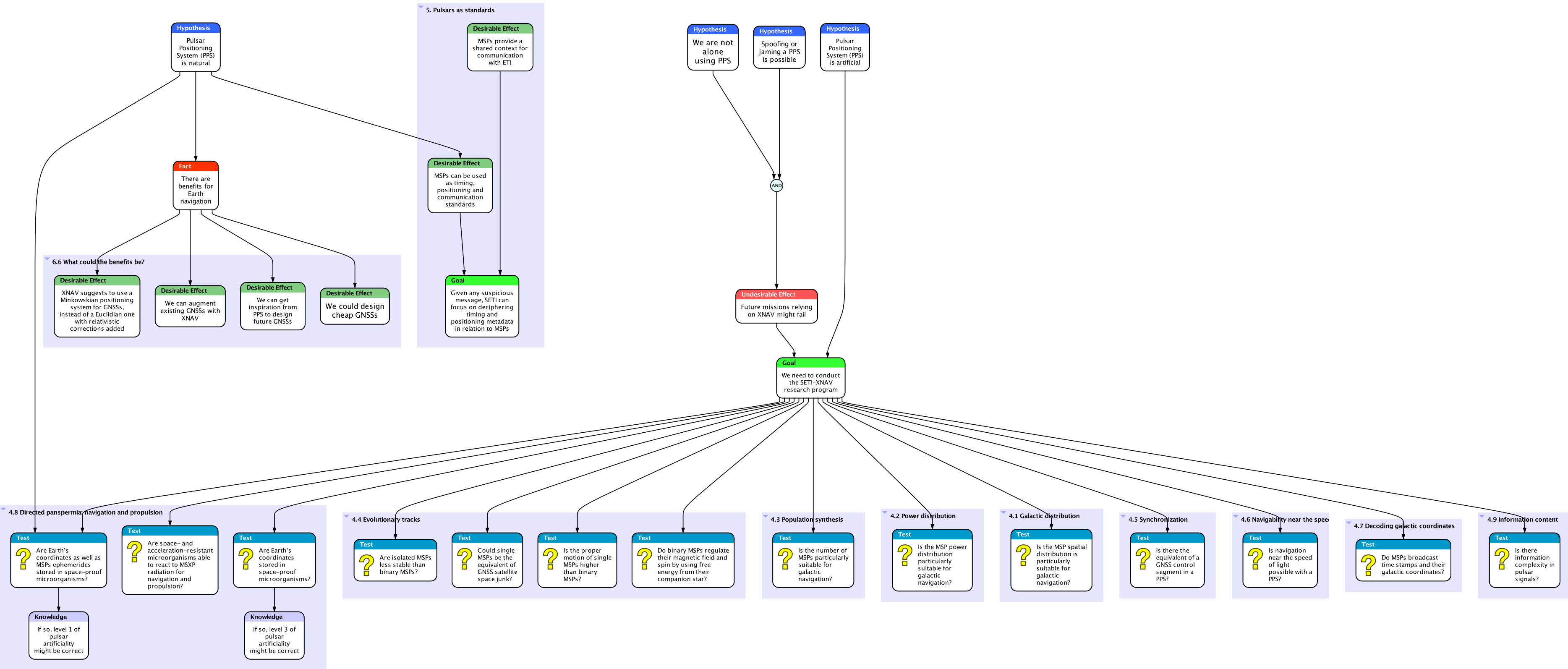